\documentclass[]{interact}

\usepackage{lmodern}
\usepackage{subcaption}
\usepackage{graphicx}    
\usepackage{url}
\usepackage{amsmath}
\usepackage{fontawesome}
\usepackage{orcidlink}
\usepackage{hyperref}
\usepackage[numbers,sort&compress]{natbib}
\bibpunct[, ]{[}{]}{,}{n}{,}{,}

\def\NAT@def@citea{\def\@citea{\NAT@separator}}



\graphicspath{{Figures/}}

\begin{document}


\title{When One Sensor Fails: Tolerating Dysfunction in Multi-Sensor Prototypes}

\author{
\name{Freek~Hens\,\orcidlink{0009-0005-0405-0560}\textsuperscript{a\,\textdagger\,*}, Amirhossein Sadough\,\orcidlink{0009-0005-5647-2888}\textsuperscript{a\,\textdagger}, Aleksa~Bok\v{s}an\,\orcidlink{0009-0003-3099-4472}\textsuperscript{b\,\textdagger}, Mahyar~Shahsavari\,\orcidlink{0000-0001-7703-6835}\textsuperscript{a} and {Mohammad Mahdi}~Dehshibi\,\orcidlink{0000-0001-8112-5419}\textsuperscript{c\,**}}
\thanks{* Freek Hens, Corresponding Author, \faEnvelope~\href{mailto:freek.hens@outlook.com}{freek.hens@outlook.com}}
\thanks{** {Mohammad Mahdi}~Dehshibi, Principal Corresponding Author, \faEnvelope~\href{mailto:mohammad.dehshibi@yahoo.com}{mohammad.dehshibi@yahoo.com}}
\thanks{\textdagger~Equal contribution.}
\affil{\textsuperscript{a}Department of Machine Learning and Neural Computing, Donders Institute for Brain, Cognition and Behaviour, Radboud University, the Netherlands \newline 
\textsuperscript{b}Delft University of Technology, Delft, the Netherlands \newline 
\textsuperscript{c}Unconventional Computing Lab., University of the West of England, Bristol, U.K.}
}

\maketitle

\begin{abstract}
Surface electromyography (sEMG) sensors are widely used in human-computer interaction, yet the failure of a single sensor can compromise system usability. We propose a methodological framework for implementing a fail-safe mechanism in multi-sensor sEMG systems. Using arm sEMG recordings of rock-paper-scissors gestures, we extracted hand-crafted features and quantified class separability via the maximum Fisher discriminant ratio (FDR). A multi-layer perceptron validated our approach, consistent with prior findings and physiological evidence. Systematic sensor ablations and FDR analysis produced a ranking of crucial versus replaceable sensors. This ranking informs robust device design, sensor redundancy, and reliability in clinical and practical applications.
\end{abstract}

\begin{keywords}
Sensor Ablation, Fault Tolerance, Gesture Recognition, Wearable Devices, Class Separability
\end{keywords}

\section{Introduction}

Multi-sensor systems are the backbone of modern Human-Computer Interaction (HCI), driving applications ranging from physiological computing and wearable devices to intelligent environments~\cite{silva2015introduction,yang2024intelligent}. However, the operational brittleness of these systems to sensor failure remains a critical bottleneck, hindering their transition from controlled laboratory prototypes to robust real-world deployments~\cite{sharma2007prevalence}. A single sensor malfunctioning, whether from signal loss, hardware faults, or battery exhaustion, can have severe consequences. Such failures can silently corrupt data collection in user studies or even render interactive systems entirely unusable, jeopardising the validity of research outcomes and the reliability of critical applications such as assistive technologies, medical devices, and safety-critical human–machine interfaces~\cite{kim2017problems}.

This challenge is particularly acute in high-stakes HCI domains. In assistive technology or remote healthcare monitoring, for instance, sensor failure is not merely an inconvenience but a direct threat to user safety and well-being~\cite{schukat2016unintended}. While established engineering fields employ fault-tolerant strategies like hardware triplication or complex software-based ``digital twins"~\cite{lyons1962use,baranwal2024faulttolerant}, these heavyweight solutions are often ill-suited to the rapid, iterative, and resource-constrained nature of HCI prototyping. The community's de facto approach often involves building a complete system and training a downstream machine learning model to assess performance---a costly, time-consuming, and fundamentally reactive process that only reveals failures after significant investment.

What is missing is a lightweight, \textit{a priori} method to audit the robustness of a sensor configuration before committing to costly data collection and model development. We argue that the HCI field needs a tool analogous to the ``zero-cost" or ``training-free" proxies emerging in Neural Architecture Search (NAS), which can estimate a model's final performance without the expensive training process~\cite{lee2024aznas}. Adapting this paradigm to HCI would enable rapid, data-driven evaluation of sensor configurations, aligning perfectly with the field's user-centred and iterative design cycles. Such a proxy would allow designers and researchers to proactively evaluate questions such as: ``How resilient is my sensor design?" and ``Which sensors are most critical for the intended task?"

\begin{figure}[!h]
  \center
  \includegraphics[width=0.7\textwidth]{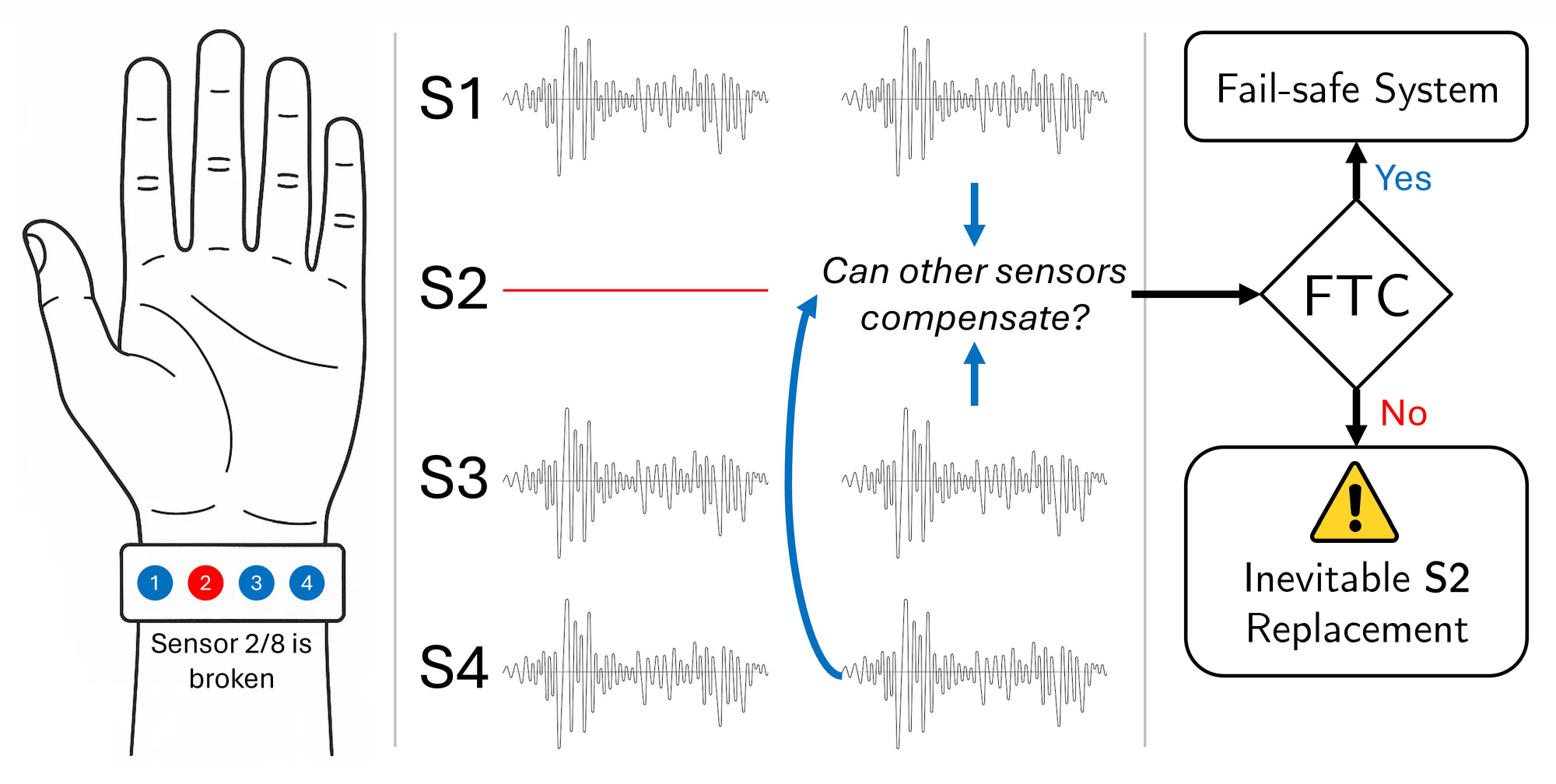}
  \caption{Fall back mechanism for sEMG-based gesture recognition.}
  \label{fig:teaser}
\end{figure}

To bridge this gap, we introduce a model-free framework for quantifying a system's Fault Tolerance Capability (see Fig.~\ref{fig:teaser}). This reliability engineering concept describes a system's robustness in the presence of component failures~\cite{gokce2013fault}. Our framework serves as a training-free proxy for performance expectations by measuring the data's inherent class separability. Using surface electromyography (sEMG)-based gesture recognition as a challenging exemplar for its notoriously low signal-to-noise ratio~\cite{wei2023multimodal}, we quantify task separability via the maximum Fisher discriminant ratio (FDR)---a robust, non-parametric metric that maximises the ratio of between-class to within-class variance~\cite{fisher1936use}. This process yields a sensor criticality ranking, distinguishing indispensable sensors from those that are replaceable.

This approach provides a generalizable, lightweight auditing tool that allows HCI developers to proactively design and validate resilient interactive systems. It enables \textit{a priori} analysis of task complexity and informs early-stage design decisions to enhance system robustness, ensuring reliability in both laboratory and real-world settings.

In summary, this paper makes the following contributions:
\begin{itemize}
    \item Introducing a generalizable, model-free framework for quantifying Fault Tolerance Capability (FTC) in multi-sensor interactive systems.
    \item Proposing a lightweight, training-free proxy for system performance and sensor criticality using the maximum Fisher discriminant ratio and systematic sensor ablation.
    \item Validating the framework on a challenging sEMG-based gesture recognition task, demonstrating its utility for \textit{a priori} task complexity analysis and robust system design.
\end{itemize}

\section{Related Work}
Our research builds on foundational work in HCI system evaluation, fault tolerance in engineering, and \textit{a priori} performance proxies in machine learning. Given the novelty of our lightweight, model-free framework for auditing the robustness in HCI prototypes, we draw on adjacent fields to highlight persistent gaps. Specifically, we trace how gaps in hardware reliability emerged in HCI's shift toward sensor-heavy systems, how prior efforts attempted to address them through reactive or heavyweight methods, and why these fell short---paving the way for our FDR-based sensor ablation approach as a proactive, training-free solution tailored to HCI's iterative design needs.

\subsection{Evaluating Interactive Systems in HCI} 
The evaluation of interactive systems has been central to HCI since its inception, evolving to address the growing complexity of multi-sensor technologies like wearables and physiological interfaces~\cite{silva2015introduction,yang2024intelligent}. Early gaps emerged from the field's focus on software usability, leaving hardware robustness underexplored. For instance, sensor failures in prototypes could invalidate user studies or deployments, although evaluations were predominantly conducted post-hoc~\cite{sharma2007prevalence,kim2017problems}.

Initial attempts involved adapting expert-oriented techniques, including heuristic evaluations~\cite{nielsen1994enhancing} and cognitive walkthroughs~\cite{lewis1997cognitive}, which facilitated cost-effective identification of design flaws in prototypes. Later, user-centred summative methods such as lab-based usability testing~\cite{rubin2008handbook} and field studies~\cite{consolvo2008activity} aimed to encompass real-world performance, including in sensor-driven human-computer interaction contexts, such as surface electromyography for gesture control~\cite{wei2023multimodal}. However, these methods remained reactive: they assess built systems and often overlook silent hardware faults, such as sEMG sensor dropout in prosthetics, which pose risks to user safety in critical areas like healthcare~\cite{schukat2016unintended,dehshibi2023pain,hens2025last}.

Surveys on evaluation methods for wearable and sensor-based systems further illustrate this limitation, revealing a predominance of post-deployment assessments that do not anticipate hardware vulnerabilities during the design phase~\cite{kim2015reliability, haresamudram2025past}. This leaves a critical gap in~\emph{a priori} hardware auditing, as HCI prototyping demands rapid iteration without costly failures. Our framework addresses this by quantifying class separability and sensor criticality in advance through the maximum Fisher discriminant ratio and ablation~\cite{fisher1936use}, complementing conventional evaluations with a data-driven proxy for robustness.

\subsection{Fault Tolerance: From Heavyweight Engineering to Lightweight HCI} 
Fault tolerance principles originated in reliability engineering to mitigate component failures in critical systems, a gap that widened as HCI incorporated multi-sensor setups prone to brittleness (e.g., signal loss in wearable devices~\cite{sharma2007prevalence}). Early engineering solutions employed hardware redundancy, such as Triple Modular Redundancy, where components are triplicated for voting-based error masking, successfully applied in aviation but at a high cost~\cite{lyons1962use}.

More recent efforts have shifted toward software alternatives, such as digital twins for simulating failures~\cite{lauer2024digital,baranwal2024faulttolerant} and, in HCI contexts, adaptive machine learning (ML) for sensor dropout in gesture recognition~\cite{ren2013robust}. For sEMG systems, studies have attempted to enhance resilience by retraining models on ablated data~\cite{ma2020emg,ma2020neuromorphic}, but these approaches remain reactive and require full datasets and models. Such approaches are impractical for HCI's resource-constrained prototyping, where failures like battery exhaustion can halt iterations~\cite{kim2017problems}.

Recent work on fault tolerance in multi-sensor fusion, particularly in autonomous driving and wearable health monitoring, has attempted to enhance system-level resilience through simulated faults and memristive associative learning circuits~\cite{bhardwaj2025memristive,wang2020design}. However, these approaches could not fully close the gap due to their heavyweight nature, incompatible with HCI's agile cycles. Our method addresses this by providing a model-free FTC assessment through sensor ablation, identifying critical sensors~\emph{a priori} without redundancy or simulation, thus enabling fail-safe designs in sEMG and similar applications.

\subsection{A Priori Performance Expectation: Bridging Machine Learning and HCI}
 In machine learning, the resource demands of model assessment~\cite{Dehshibi2024,ramin2012counting,sepas2012novel}created a gap for efficient proxies, inspiring ``zero-cost" methods in Neural Architecture Search (NAS)~\cite{mellor2021neural}. Conventional NAS entailed exhaustive candidate training~\cite{zoph2017neural}, but proxies, such as those analysing untrained network properties (e.g., expressivity), provided quick estimates~\cite{lee2024aznas}, thereby addressing the gap by predicting performance without training.

Associated data complexity metrics, such as the maximum Fisher discriminant ratio (FDR) for class separability~\cite{ho2002complexity, fisher1936use}, attempted to quantify task difficulty before modelling and were applied in pattern recognition but rarely to hardware. In HCI, this paradigm remains an emerging one. While some sEMG studies use separability post-hoc~\cite{roshambo2019dataset, roshambo}, they do not audit sensor configurations upfront, leaving prototypes vulnerable.

Recent advancements in zero-cost proxies, including evaluations of their robustness and evolutionary designs, have refined these tools for broader generalization~\cite{lukasik2025evaluation, huang2025evolving}. Nonetheless, these machine learning advancements have not been extensively integrated into human-computer interaction due to their focus on models over hardware. Our framework adapts them as a training-free proxy. In practice, measuring FDR under sensor ablations estimates robustness and task complexity early, bridging the gap for resilient multi-sensor designs in iterative HCI prototyping.

\section{A Framework for A Priori Robustness Auditing}

We introduce a general-purpose, model-free framework for auditing the robustness of multi-sensor interactive systems. This framework provides actionable insights into a system's Fault Tolerance Capability (FTC) during the early design and prototyping stages. Directly analysing the inherent statistical properties of sensor data enables researchers to make informed design decisions and anticipate potential failure points without building and deploying a complete system upfront. The framework is sensor-agnostic and comprises two primary stages: (1)~\textit{A Priori} Task Complexity Analysis, and (2)~Sensor Criticality and FTC Assessment.

\subsection{Data Representation in a Latent Feature Space}
The framework operates on the principle that the state of an interactive system can be represented as a point in a high-dimensional feature space, enabling generalised analysis.

Let $\mathbf{X} \in \mathbb{R}^{n \times m}$ be the raw data from a multi-sensor system, where $n$ is the number of samples and $m$ is the number of raw sensor channels. For each task category $c_i$ (e.g., a gesture), we apply a set of feature extraction functions $\phi_k: \mathbb{R}^m \rightarrow \mathbb{R}, k = 1, \dots, d$. This transforms the raw data into a more descriptive latent space. The result is a feature matrix for each category, $\mathbf{F}_{c_i} \in \mathbb{R}^{n_i \times d}$, where $n_i$ is the number of samples in $c_i$. This matrix serves as the basis for all subsequent analysis and is formally defined using Eq.~\eqref{eq:01}.

\begin{equation}
    \label{eq:01}
    \mathbf{F}_{c_i} = \begin{bmatrix}
        \phi_1(\mathbf{X}_{c_i, 1}) & \phi_2(\mathbf{X}_{c_i, 1}) & \cdots & \phi_d(\mathbf{X}_{c_i, 1}) \\
        \phi_1(\mathbf{X}_{c_i, 2}) & \phi_2(\mathbf{X}_{c_i, 2}) & \cdots & \phi_d(\mathbf{X}_{c_i, 2}) \\
        \vdots & \vdots & \ddots & \vdots \\
        \phi_1(\mathbf{X}_{c_i, n_i}) & \phi_2(\mathbf{X}_{c_i, n_i}) & \cdots & \phi_d(\mathbf{X}_{c_i, n_i})
    \end{bmatrix},
\end{equation}
where $\mathbf{X}_{c_i, j}$ is the $j$-th sample from category $c_i$.

\subsection{Core Analytic: Quantifying Distributional Separability}

The analytical core of our framework is a set of metrics, drawn from the data complexity literature~\cite{ho2002complexity}, that quantify the separability between two distributions in a feature space. These metrics are applied in a one-vs-rest manner, comparing a target distribution (e.g., a specific class or system state) against a reference distribution (e.g., all other classes or a baseline state).

\subsubsection{Maximum Fisher Discriminant Ratio (F1)}

This metric, also known as the maximum Fisher Discriminant Ratio (FDR), identifies the single most discriminative feature for separating a target distribution ($c_i$) from a reference distribution (``rest"). For each feature $k$, the ratio is defined using Eq.~\eqref{eq:02}.

\begin{equation}
\label{eq:02}
    \text{F1}_k(c_i) = \frac{(\mu_{i,k} - \mu_{\text{rest},k})^2}{\sigma_{i,k}^2 + \sigma_{\text{rest},k}^2},
\end{equation}
where $\mu_{i,k}$ and $\sigma_{i,k}^2$ are the mean and variance of feature $k$ for the target distribution $c_i$, with analogous definitions for the ``rest" distribution. The final score is the maximum value across all features, $\text{F1}(c_i) = \max_{k} \text{F1}_k(c_i)$. Higher F1 values indicate better separability along at least one feature dimension.

\subsubsection{Volume of Overlapping Region (F2)}

This metric quantifies the geometric overlap of the feature distributions. For each feature dimension $k$, we first define the minimum and maximum feature values for both the target ($c_i$) and reference (``rest") distributions, see Eq.~\eqref{eq:03}.

\begin{align}
\label{eq:03}
    \min_{i,k} &= \min_{x \in c_i} F_{x,k}, \quad \max_{i,k} = \max_{x \in c_i} F_{x,k}, \nonumber \\
    \min_{\text{rest},k} &= \min_{x \notin c_i} F_{x,k}, \quad \max_{\text{rest},k} = \max_{x \notin c_i} F_{x,k}.
\end{align}

The length of the overlapping segment ($\text{overlap}_k$) and the total span of values ($\text{range}_k$) for dimension $k$ are then computed using Eq.~\eqref{eq:04}.

\begin{align}
\label{eq:04}
    \text{overlap}_k &= \max\left(0, \min(\max_{i,k}, \max_{\text{rest},k}) - \max(\min_{i,k}, \min_{\text{rest},k})\right), \nonumber \\
    \text{range}_k &= \max(\max_{i,k}, \max_{\text{rest},k}) - \min(\min_{i,k}, \min_{\text{rest},k}).
\end{align}

The F2 metric is the product of these overlap ratios across all dimensions, representing the volume of the overlapping hyper-rectangle. F2 is formally defined using Eq.~\eqref{eq:05}.

\begin{equation}
\label{eq:05}
    \text{F2}(c_i) = \prod_{k=1}^d \frac{\text{overlap}_k}{\text{range}_k} \quad (\text{where } \text{range}_k > 0).
\end{equation}
where a lower F2 value signifies less overlap and thus better separability.

\subsubsection{Maximum Individual Feature Efficiency (F3)}
This metric complements F1 by identifying the single feature that provides the ``cleanest" separation, defined as the largest non-overlapping portion of the distributions (see Eq.~\eqref{eq:06}). Higher F3 values indicate better separability.

\begin{equation}
\label{eq:06}
    \mathrm{F3}(c_i) = \max_{k=1,\dots,d} \left(1 - \frac{\mathrm{overlap}_k}{\mathrm{range}_k}\right).
\end{equation}

\subsection{Stage 1: A Priori Task Complexity Analysis}
\label{sec:stage1}
\begin{figure}[!ht]
  \centering
  \begin{subfigure}{0.4\textwidth}
    \includegraphics[width=\textwidth]{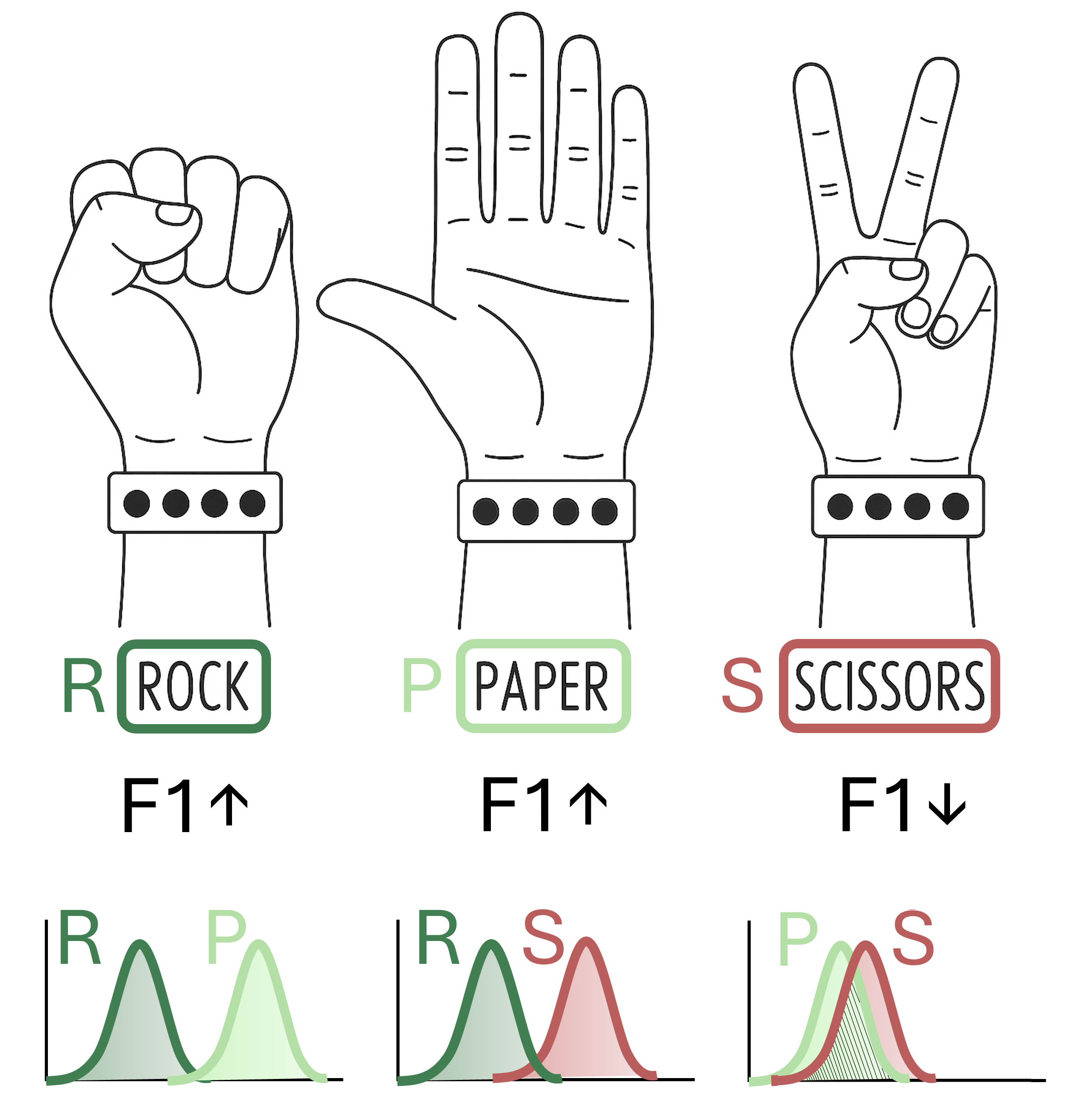}
    \caption{}
    \label{fig:framework_concept}
  \end{subfigure}
  \hspace{0.5cm}
  \begin{subfigure}{0.4\textwidth}
    \includegraphics[width=\textwidth]{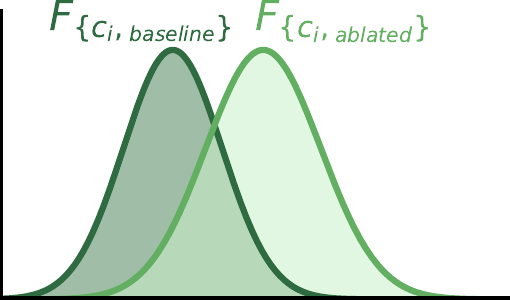}
    \caption{}
    \label{fig:baseline_ablated_distribution}
  \end{subfigure}
  \caption{(a) Conceptual illustration of class separability. The figure visualises the varying degrees of distributional overlap for three gesture classes. The simplified 1D distributions at the bottom show that `Rock' and `Paper' are well-separated (low overlap, high separability), while `Paper' and `Scissors' have a large overlapping region (shaded), indicating they are inherently more difficult to distinguish from each other based on the sensor data. (b) Conceptual illustration of distributional shift. This figure illustrates how sensor ablation affects the distribution of the baseline, i.e., the intact sensors.}
  \label{fig:framework_main}
\end{figure}

In this stage, we assess the inherent difficulty of distinguishing between different task categories. We apply the separability metrics by treating one task class as the target distribution and another (or all others) as the reference distribution. For instance, to quantify the difficulty of separating `Rock' from `Paper', we compute $\text{F2}\left(\mathbf{F}_{\text{rock}}, \mathbf{F}_{\text{paper}}\right)$. A low separability score (e.g., a high F2 value) indicates significant distributional overlap, predicting that the system will struggle to distinguish these two tasks. This concept is illustrated in Fig.~\ref{fig:framework_concept}, where the minimal overlap between the ‘Rock’ and ‘Paper’ distributions indicates an easily separable task, whereas the substantial overlap between the ‘Paper’ and ‘Scissors’ distributions suggests a considerably more challenging one. 
This provides an early diagnostic for designers to refine interactions that may be ambiguous.

\subsection{Stage 2: Sensor Criticality and FTC Assessment}
\label{sec:stage2}

In this stage, we audit the system's robustness to sensor failure. Here, we repurpose the same separability metrics to quantify each sensor's informational contribution. First, a baseline latent space distribution is established for a given class $c_i$ using all functional sensors, yielding $\mathbf{F}_{\{c_i,~\text{baseline}\}}$. Then, a sensor failure is simulated by nullifying that sensor's data stream before feature extraction, resulting in an ablated distribution, $\mathbf{F}_{\{c_i,~\text{ablated}\}}$. The magnitude of the distributional shift is quantified by computing the separability between these two states, e.g., $\text{F1}\left(\mathbf{F}_{\{c_i,~\text{baseline}\}}, \mathbf{F}_{\{c_i,~\text{ablated}\}}\right)$. This concept is illustrated in Fig.~\ref{fig:baseline_ablated_distribution}, where

\begin{itemize}
    \item a large shift (high F1/F3, low F2) implies the sensor's absence fundamentally alters the data's structure. Such a sensor is deemed \textit{indispensable} for that task.
    \item a minimal shift (low F1/F3, high F2) implies that other sensors sufficiently capture the necessary information, rendering this sensor redundant or \textit{replaceable}.
\end{itemize}

This analysis, extendable to combinatorial ablations (pairs, triplets), yields a sensor criticality ranking and identifies the system's tolerance thresholds, providing an actionable blueprint for designing robust, fail-safe systems.

\section{Case Study: Auditing Robustness in an sEMG-based Gesture Interface} 

To demonstrate and validate our framework in a realistic HCI scenario, we apply it to a publicly available sEMG hand-gesture recognition dataset~\cite{garg2020signals}. This case study serves as a challenging exemplar for three reasons: (1) it employs a multi-sensor sEMG system, a common but notoriously noisy modality; (2) the gesture set includes classes known to be difficult to distinguish, providing a robust test for our task complexity analysis; and (3) the original publication provides baseline classification results, allowing for direct validation of our framework's predictive power.

\subsection{Case Study Context: The Roshambo Dataset}

\begin{figure}[!ht]
    \centering
    \includegraphics[width=0.5\textwidth]{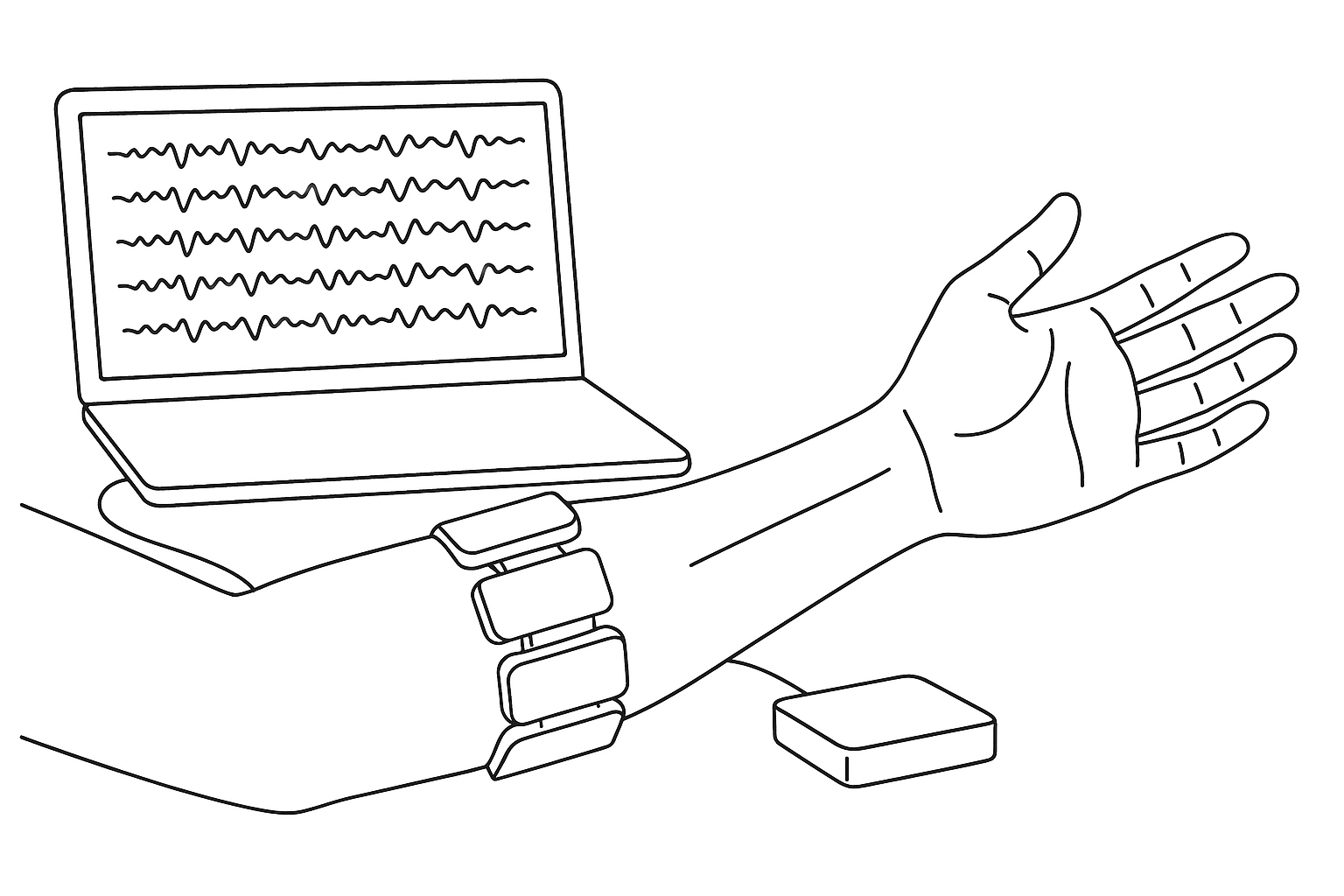}
    \caption{Conceptual illustration of the data collection setup using the Myo armband with eight sEMG sensors, based on the description in Garg et al.~\cite{garg2020signals}. The original dataset does not specify the anatomical mapping for each sensor index.} 
    \label{fig:myoband}
\end{figure}

The dataset contains recordings from ten participants using a Myo armband, which has eight sEMG sensors evenly distributed around the forearm and operates at a sampling frequency
of 200~Hz~\cite{roshambo2019dataset} (see Fig.~\ref{fig:myoband}). 

Participants performed three gestures (i.e., rock, paper, and scissors) alongside a control state representing the arm and hand in a resting position. The protocol consisted of three sessions for each participant, with five trials lasting 3 seconds each for each gesture, resulting in a total of 450 trials. To ensure accurate benchmarking, the initial and final 600 ms of each trial were trimmed to focus on the steady-state portion of the gestures. For our analysis, we employed a standard sliding-window approach (400 timestamps, or 2 seconds, with 50\% overlap) on the processed signals. This methodology produced 296 uniform samples per class, each with dimensions of $8 \times 400$ (sensors~$\times$~timestamps).

\subsection{Audit Configuration}

\noindent\textbf{Feature Selection.} Following established practice in sEMG analysis~\cite{Phinyomark2012PartialHand}, we extracted nine general-purpose features for each sensor. This set includes \textit{Shannon Entropy}, \textit{Sample Entropy}, \textit{Zero Crossings}, \textit{Waveform Length}, \textit{Root Mean Square (RMS)}, \textit{Slope Sign Changes}, \textit{Median Frequency}, \textit{Wavelet Energy}, and \textit{Fractal Dimension}. This comprehensive set was deliberately chosen to be modality-agnostic, capturing a broad range of a signal's temporal, spectral, and complexity characteristics. This approach ensures our audit's findings are not dependent on domain-specific feature engineering. The extraction process transformed each $8 \times 400$ sample into a 72-dimensional feature vector (8 sensors~$\times$~9 features).\newline

\noindent\textbf{Validation Oracle.} To validate our framework's \textit{a priori} predictions, we implemented a simple Multi-Layer Perceptron (MLP) classifier. The goal was not to achieve state-of-the-art accuracy, but to use a standard classifier as an ``oracle" to confirm whether the separability issues predicted by our framework manifest in a typical learning model. We adopted a one-vs-one binary classification design (e.g., `paper' vs `scissors') rather than a single multiclass model. This approach provides a fine-grained ``magnifying glass" to directly test the pairwise separability predictions from our framework, which would be obscured in an aggregate multiclass accuracy score. Performance was measured with the Matthew's Correlation Coefficient (MCC)~\cite{matthews1975comparison}, a robust metric for binary classification.

\subsection{Results and Implications for Design}

\subsubsection{Stage 1: A Priori Prediction of Task Difficulty} 

\begin{figure}[!ht]
    \centering
    \begin{subfigure}[t]{0.35\textwidth}
        \centering
        \includegraphics[width=\textwidth]{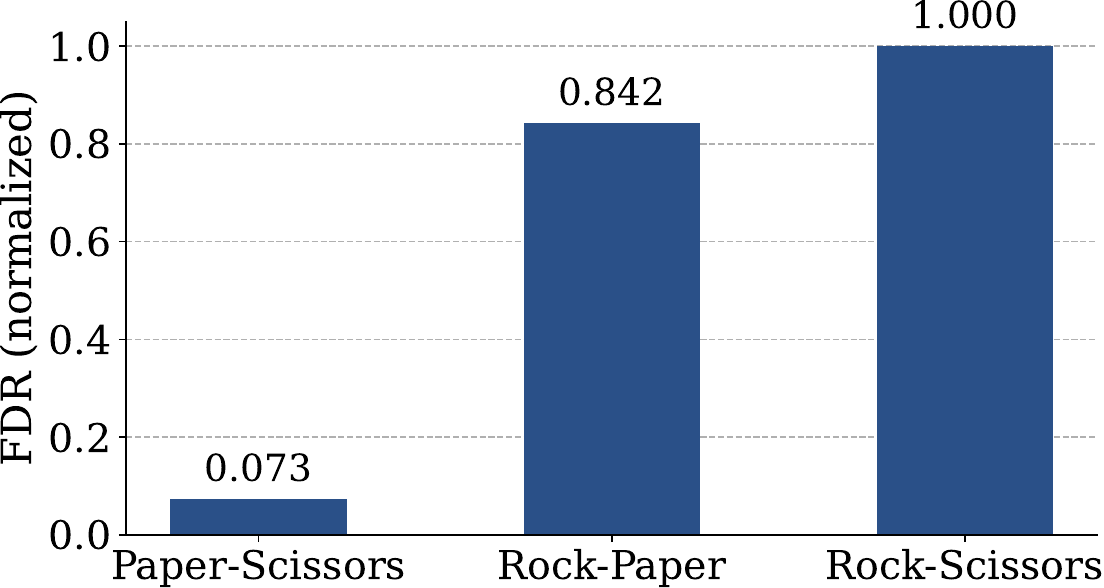}
        \caption{}
    \end{subfigure}
    \hspace{0.5cm}
    \begin{subfigure}[t]{0.355\textwidth}
        \centering
        \includegraphics[width=\textwidth]{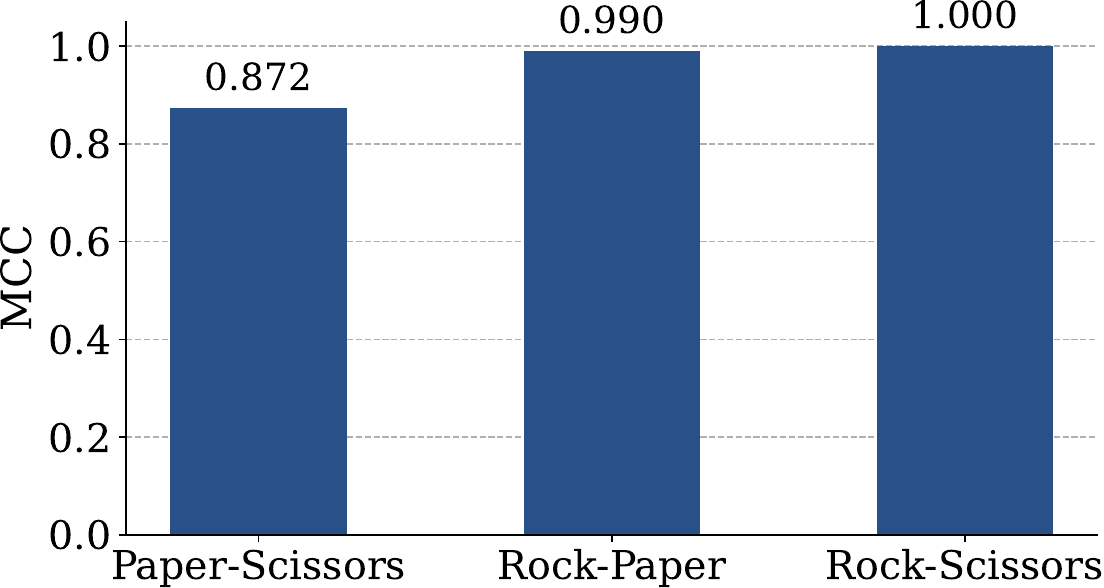}
        \caption{}
        \end{subfigure}
    \caption{Stage 1 Results: (a) Our framework's model-free analysis predicts that the `paper' vs `scissors' task is over $10\times$ more difficult (lower FDR) than the other pairs. (b) This prediction is validated by an MLP classifier, which achieves a demonstrably lower MCC score on the same pair.}
    \label{fig:results_stage1}
\end{figure}

Our framework's audit immediately and correctly identified the inherent difficulty in the gesture set. The analysis was conducted using the one-vs-one Fisher's Discriminant Ratio (FDR), a direct measure of class separability equivalent to the F1 metric in a binary context.
As shown in Fig.~\ref{fig:results_stage1}(a), the `paper' vs `scissors' pair yielded a considerably low normalised FDR of 0.073, indicating severe overlap in their feature spaces. In stark contrast, the `rock' vs `paper' and `rock' vs `scissors' pairs produced high normalised FDR scores of 0.842 and 1.00, respectively, indicating that they would be easily distinguishable.

This model-free prediction was directly validated by the MLP's performance (Figure~\ref{fig:results_stage1}(b)). The `paper' vs `scissors' classifier achieved a significantly lower MCC of 0.90 (corresponding to 95\% accuracy), whereas the other two pairs achieved near-perfect MCC scores approaching 1.0. The near-perfect validation scores are not an indicator of overfitting, but rather a confirmation of the extreme class separability for those pairs, as predicted by our framework's high FDR scores. Our findings are in complete agreement with the confusion matrices reported in the original Roshambo paper~\cite{garg2020signals}. Additionally, the physiological basis for this difficulty is well-established: `rock' primarily involves a distinct set of flexor muscles, while both `paper' and `scissors' rely on overlapping extensor muscles in the forearm~\cite{furui2019myoelectric}. Our framework successfully identified this underlying physiological ambiguity directly from the signal data, validating its use as a powerful proxy for task complexity.

\subsubsection{Stage 2: Sensor Criticality and Neighbour Compensation Analysis} 

\begin{figure}[!h]
    \centering
    \begin{subfigure}[t]{0.32\textwidth}
        \centering
        \includegraphics[width=\textwidth]{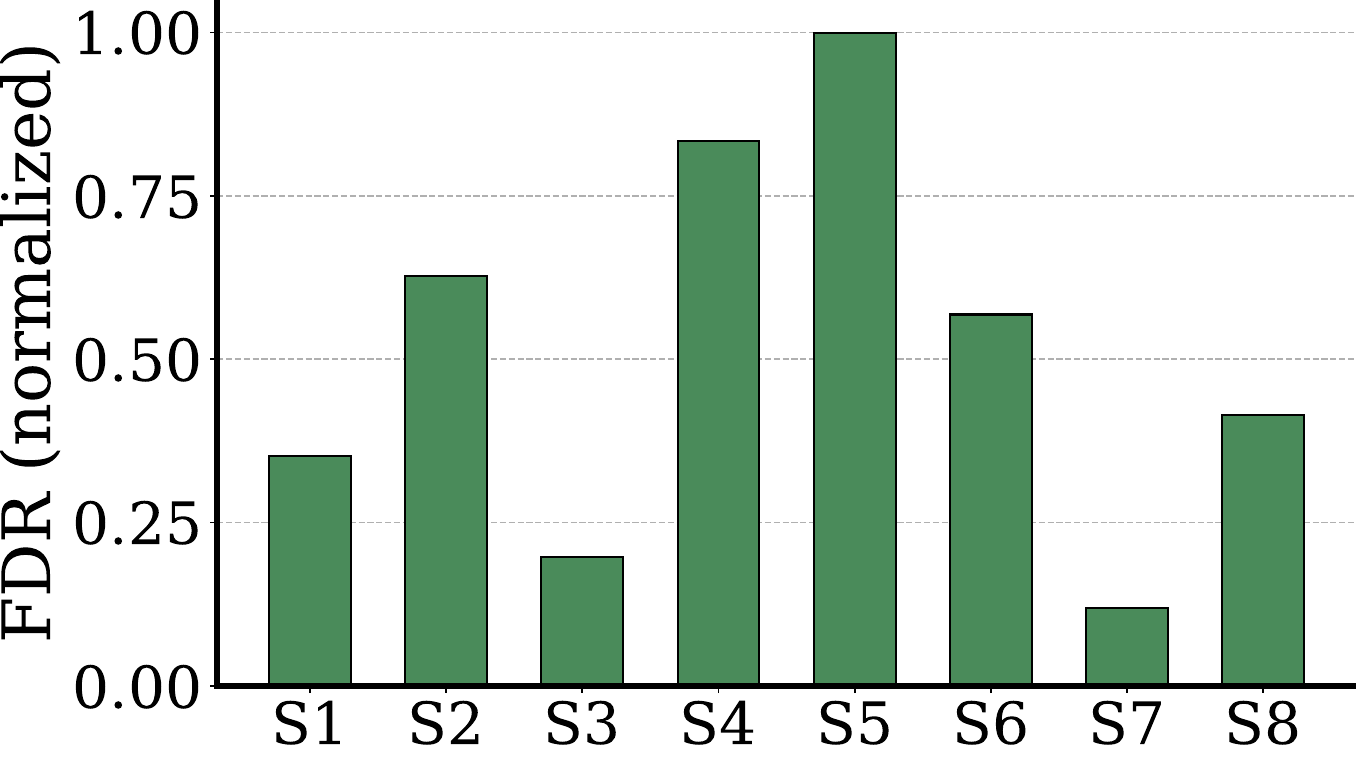}
        \caption{Rock}
    \end{subfigure}
    \hfill
    \begin{subfigure}[t]{0.32\textwidth}
        \centering
        \includegraphics[width=\textwidth]{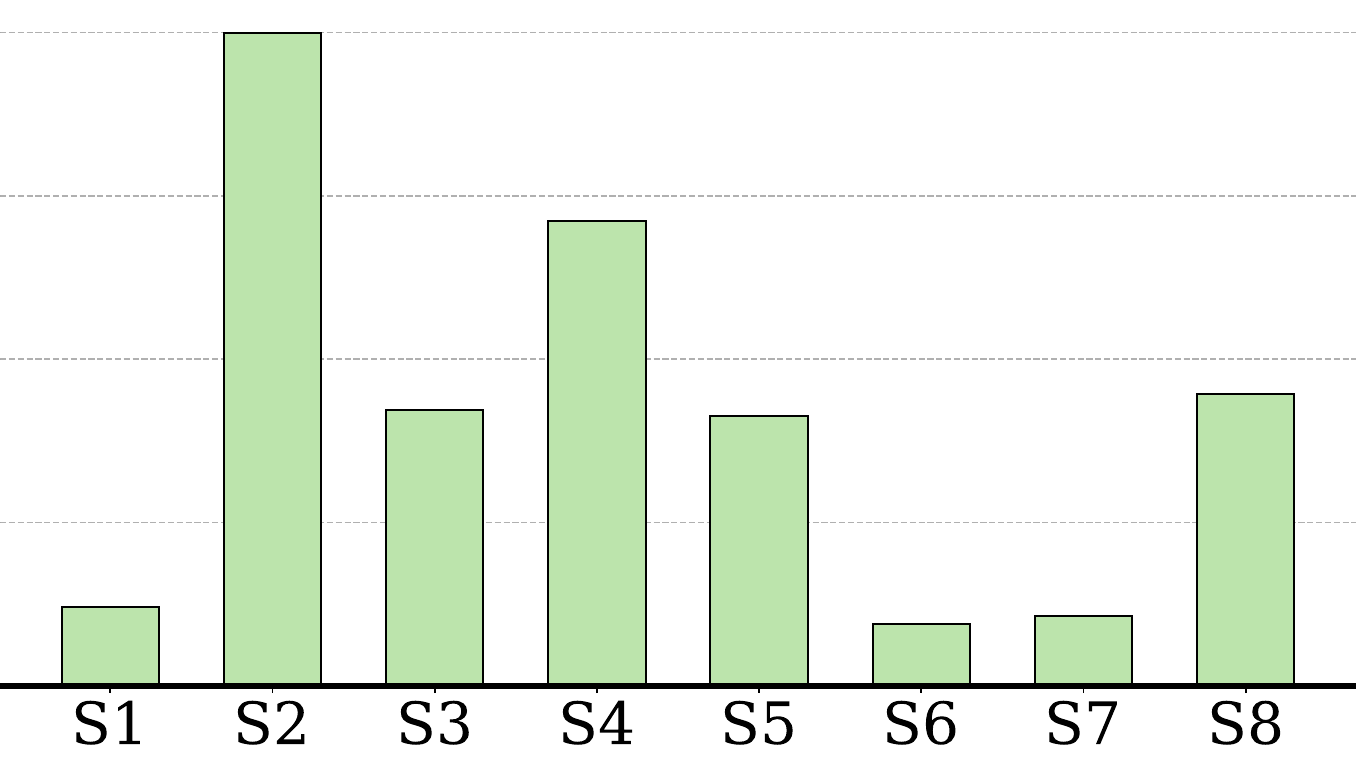}
        \caption{Paper}
    \end{subfigure}
    \hfill
    \begin{subfigure}[t]{0.32\textwidth}
        \centering
        \includegraphics[width=\textwidth]{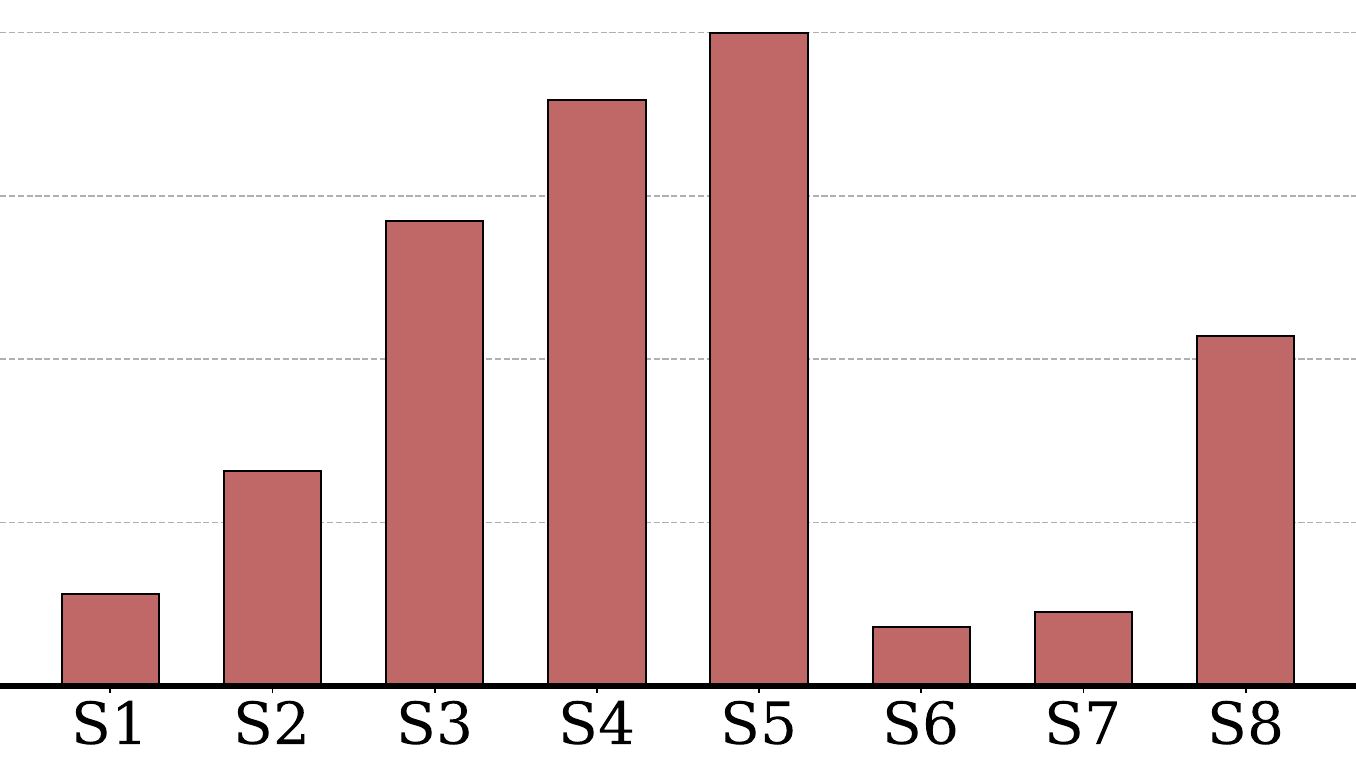}
        \caption{Scissors}
    \end{subfigure}
    \caption{Stage 2 Results: Per-class sensor criticality. A larger bar indicates a more critical sensor for that specific gesture. The analysis identifies both indispensable sensors (e.g., Sensor 2 for Paper) and redundant ones (e.g., Sensors 6, 7).}
    \label{fig:sensor_ablation}
\end{figure}

The sensor ablation audit revealed that sensor importance is highly task-dependent. Figure~\ref{fig:sensor_ablation} shows the distributional shift (FDR) caused by ablating each sensor for each gesture.

\noindent\textbf{Neighbour Compensation Analysis.}  By analysing the scores in the context of the armband's circular layout, we can infer regional redundancy. For the `paper' gesture, Sensor 2 is highly critical. Its immediate neighbours, Sensor 1 and Sensor 3, are far less critical. This suggests the information captured by the sensor at position 2 is highly localised and \textit{cannot} be effectively compensated for by adjacent sensors, making it a critical single point of failure for this gesture. Conversely, for `scissors', the low criticality of adjacent sensors 6 and 7 suggests a region of high redundancy, where the failure of one is likely to be tolerated. \newline

\noindent\textbf{Implications for Interaction Design.} A designer armed with this audit during prototyping could: 
\begin{enumerate} 
\item \textbf{Reinforce Critical Components:} Knowing that the sensors at positions 2, 3, 4, and 5 are frequently critical, they would ensure these locations on their own prototype have robust physical placement and perform targeted data integrity checks on their signal streams. 
\item \textbf{Implement Graceful Degradation:} If the sensor at position 2 fails, the system's software could be designed to alert the user that `paper' gesture recognition may be unreliable, while other gestures remain fully functional. 
\item \textbf{Optimise for Efficiency:} Sensors 6 and 7 consistently produce minimal distributional shifts across all gestures, indicating high redundancy. A designer could confidently experiment with removing these sensors in a future iteration to reduce hardware costs, complexity, and power consumption, with a predictable, minimal impact on performance. 
\end{enumerate} 

This case study demonstrates the framework's ability to go beyond simple performance metrics, providing deep, actionable insights to build more robust, reliable, and efficient interactive systems \textit{a priori}.

\section{Discussion}
In this paper, we introduced a model-free framework for auditing the robustness of multi-sensor interactive systems \textit{a priori}. The central contribution is a lightweight, data-driven method that provides HCI researchers and practitioners with early insights into a system's potential failure points, moving beyond the conventional `build-then-test' paradigm. Our goal was to enable a more proactive design cycle where decisions about system reliability can be made before committing to costly data collection and model training.

Our case study on the sEMG-based Roshambo dataset served to validate the framework's core hypotheses. The principal findings were twofold. First, the framework's task complexity analysis correctly predicted that the `paper' and `scissors' gestures were inherently difficult to distinguish, a prediction that was subsequently confirmed by our validation MLP and the findings of the original dataset authors~\cite{garg2020signals}. This demonstrates that quantifying class separability can serve as a powerful and accurate proxy for downstream performance. Second, the sensor criticality audit provided a granular, task-dependent ranking of each sensor's importance, identifying both indispensable components and redundant ones.

\subsection{Implications for HCI Design and Practice}
The true value of this framework lies in its ability to inform concrete design decisions. An \textit{a priori} robustness audit empowers HCI teams in several ways:

\noindent\textbf{1. Proactively Shaping Interaction Design.} The framework can act as an early-warning system for interactional ambiguity. If the audit reveals high distributional overlap between two intended gestures or actions, designers can intervene early. They might choose to redesign the interactions to be more physically distinct, provide clearer user feedback, or even prune an ambiguous action from the set entirely, thereby preventing user frustration and improving system accuracy before a single line of application code is written.

\noindent\textbf{2. Informing Hardware Prototyping and Sensor Selection.} The sensor criticality ranking provides an empirical basis for hardware decisions. Teams can use this audit to justify allocating resources to higher-quality sensors for positions identified as critical. Conversely, they can explore cost and complexity reductions by removing sensors identified as consistently redundant. This data-driven approach moves hardware design beyond intuition, allowing for a more deliberate and defensible prototyping process.

\noindent\textbf{3. Designing for Graceful Degradation.} Silent failures are a notorious problem in interactive systems. Our framework provides the necessary map to design intelligent, fail-aware systems. When a sensor identified by the audit as critical begins to drift or fail, the system can be designed to respond gracefully. Rather than failing completely, it could trigger a targeted recalibration, temporarily disable the specific interactions that rely on that sensor, or notify the user that certain functions may be unreliable, thus preserving the usability of the rest of the system.

\subsection{Limitations and Future Work}
Our work provides a strong foundation, but it has limitations that encourage future research contributions. Our validation was conducted on a single, albeit challenging, sensing modality (sEMG) and dataset. Future work should validate the framework's generalizability across other modalities, such as accelerometers, electroencephalography (EEG), and capacitive sensing, as well as on various interactive tasks.

The current analysis primarily focused on single-sensor failures. While we suggest that this can be extended to combinatorial failures, a systematic investigation into multi-sensor failure regimes is a critical next step in fully understanding a system's tolerance thresholds. Furthermore, our audit was performed on a static feature space. An interesting line of future inquiry would be to examine whether the same auditability is preserved in dynamically learned feature representations from deep learning models. Finally, embedding this audit as a real-time, online component within an interactive system could provide continuous health checks, adapt to sensor drift, and provide dynamic reliability assurances to the user.

\section{Conclusion}
This paper presented a general-purpose, model-free framework for the early auditing of fault tolerance in multi-sensor HCI systems. By providing a training-free proxy for system performance, our method allows designers to identify interactional ambiguities and critical hardware components during the initial stages of prototyping. The case study demonstrated that this approach can accurately predict system vulnerabilities and yield actionable insights for building more robust, reliable, and efficient interactive systems. Ultimately, this work contributes a practical \textit{pre-mortem} tool for the HCI community, aiming to reduce wasted iteration and surface system failure modes when they are cheapest and easiest to correct at the very beginning.

\section*{Acknowledgement}
The authors declare no competing interests.

\bibliographystyle{plain}
\bibliography{refs}

\end{document}